\documentclass[aps,prl,showkeys,twocolumn,nofootinbib,superscriptaddress]{revtex4-1}
\usepackage{natbib}
\usepackage{amssymb}
\usepackage{color}
\usepackage{amsfonts}
\usepackage{textcomp}

\usepackage{float}
\usepackage[caption = false]{subfig}
\usepackage{graphicx}

\begin{document}
\title{Towards the insulator-to-metal transition at the surface of ion-gated nanocrystalline diamond films}
\author{Erik Piatti}
\author{Francesco Galanti}
\affiliation{Department of Applied Science and Technology, Politecnico di Torino, 10129 Torino, Italy}
\author{Giulia Pippione}
\affiliation{Department of Physics and ``NIS'' Centre, University of Torino, 10125 Torino, Italy}
\affiliation{Institute of Electron Devices and Circuits, Ulm University, 89069 Ulm, Germany}
\author{Alberto Pasquarelli}
\affiliation{Institute of Electron Devices and Circuits, Ulm University, 89069 Ulm, Germany}
\author{Renato S. Gonnelli}
\affiliation{Department of Applied Science and Technology, Politecnico di Torino, 10129 Torino, Italy}
\begin{abstract}
Hole doping can control the conductivity of diamond either through boron substitution, or carrier accumulation in a field-effect transistor. In this work, we combine the two methods to investigate the insulator-to-metal transition at the surface of nanocrystalline diamond films. The finite boron doping strongly increases the maximum hole density which can be induced electrostatically with respect to intrinsic diamond. The ionic gate pushes the conductivity of the film surface away from the variable-range hopping regime and into the quantum critical regime. However, the combination of the strong intrinsic surface disorder due to a non-negligible surface roughness, and the introduction of extra scattering centers by the ionic gate, prevents the surface accumulation layer to reach the metallic regime.
\end{abstract} 
\maketitle
%
Diamond is a promising material for electronic applications thanks to its wide band gap, large carrier mobility, high thermal conductivity and good electrochemical properties \cite{PanBook1995}. When boron (B) is substituted to carbon in the lattice structure, diamond becomes a p-type conductor and eventually a superconductor with a critical temperature $\sim 4$ K \cite{EkimovNature2004,BustarretPSS2008}, which can be further enhanced by optimizing the growth process \cite{TakanoAPL2004,OkazakiAPL2015}. Diamond is also considered a candidate for high-temperature superconductivity (SC) \cite{BoeriPRL2004,LeePRL2004,BoeriJPCS2005,GiustinoPRL2007,NakamuraPRB2013} thanks to its very large Debye temperature $\approx 2000\,\mathrm{K}$, but its attainability is hampered by disorder and the maximum B solubility \cite{BustarretPSS2008,TakanoAPL2004,OkazakiAPL2015}. Charge carriers can also be introduced in an insulator via the electric field effect in the form of a 2-dimensional (2D) surface accumulation layer, upon application of a transverse electric field. The electric-double-layer (EDL) transistor is a particularly effective field-effect architecture, thanks to the large electric field that develops at a solid/electrolyte interface when a voltage drop is applied across it \cite{UenoReview2014}. The electric transport properties of diamond-based EDL transistors have been extensively studied \cite{DankerlPRL2011,DankerlAPL2012,YamaguchiJPSJ2013,TakahidePRB2014,TakahidePRB2016,AkhgarNanoLett2016},  but no evidence was reported for either SC or good metallic behavior. This can be associated with the relatively modest charge carrier densities achieved in these works on insulating single crystals and epitaxial thin films ($\lesssim 7\cdot 10^{13}$ cm$^{-2}$ \cite{DankerlPRL2011,DankerlAPL2012,YamaguchiJPSJ2013,TakahidePRB2014,TakahidePRB2016,AkhgarNanoLett2016}) with respect to the values typically obtained in conducting systems ($\simeq10^{14}-10^{15}$ cm$^{-2}$ \cite{DagheroPRL2012,TortelloApsusc2013,PiattiPRB2017,PiattiArXiv2018,GonnelliSciRep2015,PiattiJSNM2016,LiNature2016}).

In this work, we focus on nanocrystalline diamond films and employ ionic gating to tune their surface conductivity via electrostatic hole doping. The films are slightly B-doped in order to provide a finite carrier density even in absence of an applied gate bias, and we show that this enhances the maximum induced hole density by a factor $\sim 3$ with respect to earlier reports. Despite this improvement, we don't observe SC or good metallic behavior down to $\approx 3$ K. Instead, charge transport in the accumulation layer (AL) at the film surface moves from the variable-range hopping regime at low hole doping, into the quantum critical (QC) regime at high hole doping, never crossing into the metallic side of the insulator-to-metal transition (MIT). We propose that this frustrated behavior is associated to a combination of non-negligible surface roughness and additional scattering centers introduced by the ions in the EDL.

\begin{figure}
\begin{center}
\includegraphics[keepaspectratio,height=115mm]{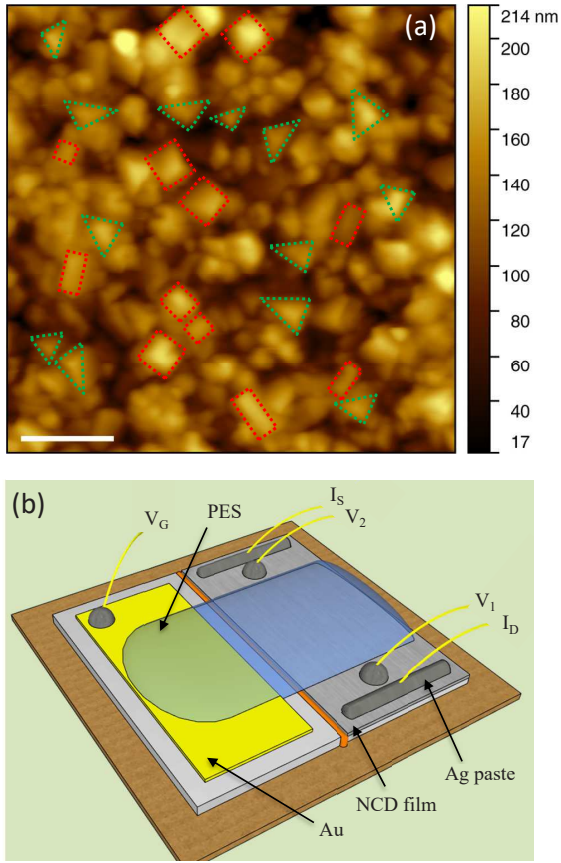}
\end{center}
\caption{(a) Representative AFM topography map acquired in contact mode of the B-doped NCD thin film before device fabrication. Red and green dashed lines outline selected square and triangular grains. Scale bar is $1\,\mathrm{\mu m}$. (b) Sketch of a complete EDLT device with the electrical connections required for four-wire resistance and gating measurements.}
\label{fig:fabrication} 
\end{figure}

Nanocrystalline diamond (NCD) films were grown by a two-step Chemical Vapor Deposition (CVD) process on AF32eco high-temperature glass substrates. A detailed description of the growth process has been reported elsewhere \cite{PippionePSS2017}. The films consisted of an underlying intrinsic NCD layer ($\simeq 1\,\mathrm{\mu m}$) with a B-doped NCD layer ($\simeq 300$ nm) on top, as determined by Scanning Electron Microscopy \cite{PippionePSS2017}. 
Fig.\ref{fig:fabrication}a shows a representative topography image acquired via Atomic Force Microscopy (AFM) in contact mode. The sample is polycrystalline with an average lateral grain size of $222\pm15$ nm. The largest grains typically grew in either rectangular (dashed red lines) or triangular (dashed green lines) shapes, which can be associated to $[100]$- and $[111]$-oriented facets respectively \cite{UshizawaDRM1998}. The average mean square roughness is \mbox{$S_q=32.3\pm0.5$ nm}, and the rugosity -- defined as the ratio between the real and projected areas of the surface -- is $f_r=1.08\pm0.02$. The amount of B doping was estimated by Hall effect experiments performed in the van der Pauw configuration at room temperature, applying a magnetic field up to $B = \pm 0.7$ T with a calibrated permanent magnet. The resulting free holonic density was $n_h = 3.2 \pm 0.5 \cdot 10^{20}$ e$^{+}$/cm$^{3}$, corresponding to a B-doping concentration of $0.36 \pm 0.06$ \%, and consistent with Ref.\cite{PippionePSS2017}. 

Fig.\ref{fig:fabrication}b shows a sketch of the EDL device, which was obtained by dicing the NCD film into smaller rectangles ($\sim 0.7\times0.3$ cm$^2$), and drop-casting a small amount of conductive silver paste to realize electrical contacts for source ($S$), drain ($D$), and four-wire voltage probes ($V_1$ and $V_2$). The EDL transistor was completed by drop-casting the liquid precursor to the cross-linked polymer electrolyte system (PES) on top of the active channel of the device -- defined by the two voltage contacts -- and the gold side gate ($V_G$). The PES consisted of a mixture of BEMA dimethacrylate oligomer and EMIM-TFSI ionic liquid in 2:8 ratio along with 3 wt\% of photo initiator, and was subsequently UV-cured in a dry room.


We measured the four-wire resistance $R$ as a function of temperature $T$ in the high-vacuum chamber ($p\lesssim 10^{-5}$ mbar) of a pulse-tube cryocooler with a base $T\simeq 3$ K, by sourcing a probe DC current $I_{DS}$ of few $\mathrm{\mu A}$ and measuring the longitudinal voltage drop $V_{12}$ across the active channel. Common mode offsets were eliminated with the current reversal method. We applied $V_G$ at $T\simeq240$ K, slightly above the glass transition of the PES, in order to minimize the chance of electrochemical interactions between the sample and the electrolyte. For each $V_G$ application, the induced charge density is $\Delta n_{2D} = Q_{EDL}/(f_r\cdot S)$, where $S$ is the surface area of the NCD film covered by the PES, $f_r$ is the rugosity determined via AFM, and $Q_{EDL}$ is the charge stored in the EDL as measured by double-step chronocoulometry. This is a well-established electrochemical technique \cite{InzeltBook} which has been employed to reliably determine $\Delta n_{2D}$ for a wide variety of materials \cite{DagheroPRL2012,TortelloApsusc2013,PiattiPRB2017,PiattiArXiv2018,GonnelliSciRep2015,PiattiJSNM2016}.  

\begin{figure}
\begin{center}
\includegraphics[keepaspectratio,height=115mm]{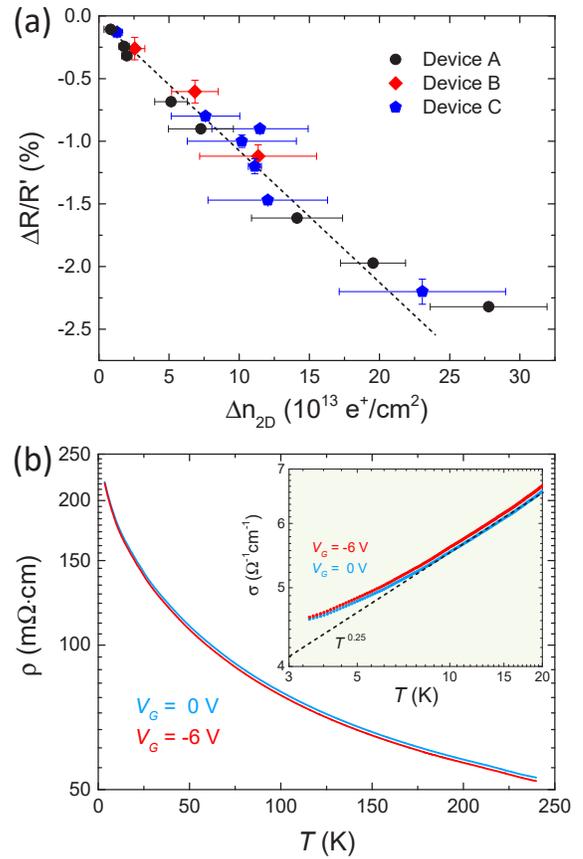}
\end{center}
\caption{
(a) Normalized resistance variation, $\Delta R/R'$, vs. induced charge density, $\Delta n_{2D}$, at $T\simeq240$ K. Different symbols refer to different EDL devices obtained from the same NCD film. Black dashed line is a guide to the eye. 
(b) Resistivity, $\rho$, vs. temperature, $T$, for Device A at $V_G = 0$ and $V_G = -6$ V, in semilogarithmic scale. Inset shows the corresponding low-$T$ conductivity in log-log scale. Dashed line is a linear fit to the range $8-20$ K.
}
\label{fig:bulk_transport} 
\end{figure}

\begin{figure*}
\begin{center}
\includegraphics[keepaspectratio,width=\textwidth]{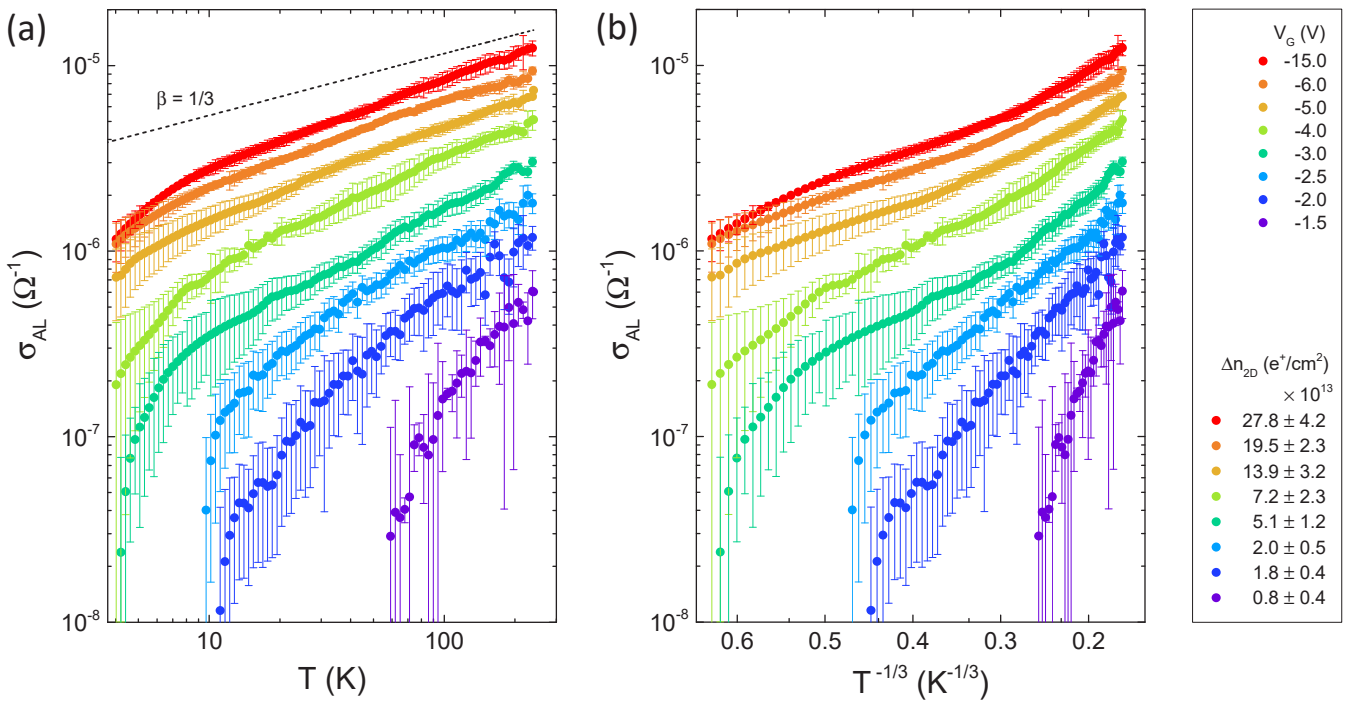}
\end{center}
\caption{
(a) Conductivity of the accumulation layer, $\sigma_{AL}$, vs. temperature $T$, in double-logarithmic scale. Dashed line indicates the scaling predicted exactly at the QC point of the MIT ($\beta=1/3$). Only a subset of the experimental data for each curve is plotted for clarity. (b) Same data of (a) plotted vs. $T^{-1/3}$. 
}
\label{fig:surface_transport}
\end{figure*}

At constant $T\simeq 240$ K, applying $V_G<0$ increased the conductivity of our devices, while applying $V_G>0$ suppressed it. All modulations were entirely reversible by simply applying $V_G = 0$. These features indicate that $V_G$ tunes the conductivity by means of purely electrostatic accumulation/depletion of holonic charge carriers in the valence band of the NCD film. In the following, we concentrate on hole accumulation  ($V_G<0$). Fig.\ref{fig:bulk_transport}a shows the normalized resistance variation \mbox{$\Delta R/R' = [R(V_G) - R(0)]/R(V_G)$} as a function of $\Delta n_{2D}$ at $T\simeq240$ K for three different NCD devices. Horizontal and vertical error bars indicate the difference in the values of $\Delta n_{2D}$ and $R$ determined between the charge and discharge of the EDL capacitor respectively. The maximum values of $\Delta n_{2D}$ are more than $3$ times larger than the ones in earlier reports \cite{DankerlPRL2011,DankerlAPL2012,YamaguchiJPSJ2013,TakahidePRB2014,TakahidePRB2016,AkhgarNanoLett2016}. The negative slope of $\Delta R/R'$ for increasing hole density indicates that charge transport is dominated by holonic carriers, consistent with the preliminary Hall measurements on the pristine film. Up to $\sim 2\cdot 10^{14}\,\mathrm{e^+/cm^2}$, $\Delta R/R'$ shows a good linear scaling with $\Delta n_{2D}$, which is expected when the gate-induced modulation to the conductivity is dominated by electrostatic charge doping \cite{DagheroPRL2012,TortelloApsusc2013,PiattiPRB2017,PiattiArXiv2018}. 


Fig.\ref{fig:bulk_transport}b shows the $T$-dependent resistivity $\rho (T)$ at $V_G = 0$ (blue curve) and \mbox{$V_G = -6$ V} (red curve) in semilogarithmic scale. In both cases, $\rho$ decreases with increasing $T$, which may suggest a semiconducting behavior. However, $\rho$ tends to saturate below $\sim 8$ K, which is the hallmark of a disordered metal \cite{McMillanPRB1981,LarkinJETP1982,HeegerPS2002}. This is best shown by plotting the $T$-dependence of $\sigma$ below $20$ K in double logarithmic scale (see the inset to Fig.\ref{fig:bulk_transport}b). Between $8$ and \mbox{$20$ K}, $\sigma(T)$ at $V_G = 0$ scales well with a power law $T^\beta$, with $\beta = 0.25\pm0.10$; below $8$ K, $\sigma$ is larger than predicted by the scaling law and shows an incipient saturation. This indicates that, even at $V_G = 0$, the bulk charge transport is in the QC regime just on the metallic side of the MIT \cite{McMillanPRB1981,LarkinJETP1982,HeegerPS2002}. This is consistent with the pristine carrier concentration determined by Hall effect ($n_h \simeq 3.2 \cdot 10^{20}$ e$^{+}$/cm$^{3}$), as the bulk MIT occurs at a slighly lower doping $\approx 2\cdot 10^{20}$ e$^{+}$/cm$^{3}$ \cite{PippionePSS2017,GononJAP1995}. Since the induction of additional charge carriers in the system ($V_G = -6$ V) reduces $\rho$/increases $\sigma$ in the entire $T$ range without an obvious change in the scaling at low $T$ (see the inset to Fig.\ref{fig:bulk_transport}b), the total $\sigma$ is completely dominated by the bulk contribution for any $V_G$.

We disentangle the charge transport of the gate-induced AL at the surface from the predominant bulk contribution through the method described in Ref.\cite{KounoArXiv2018}. For each value of $V_G$, we define the sheet conductance of the AL, $\sigma_{AL}(T)$, as the difference between the total sheet conductance of the film, $\sigma_s(T)$, at finite $V_G$ and at $V_G=0$:
\begin{equation}
\sigma_{AL}(T)|_{V_G} = \sigma_{s}(T)|_{V_G} - \sigma_{s}(T)|_{V_G=0}
\label{eq_surf}
\end{equation}
Since $\sigma_{AL}$ is small compared to $\sigma_s$, we measure $\sigma_{s}(T)|_{V_G=0}$ both before and after each $\sigma_{s}(T)|_{V_G}$. For each $V_G$ value, we then calculate $\sigma_{AL}$ in the two cases and take their average, while their difference is taken as the uncertainty. Fig.\ref{fig:surface_transport} shows the behavior of $\sigma_{AL}(T)$ with increasing negative gate bias $V_G$ and surface carrier density $\Delta n_{2D}$. 

It is immediately clear that not all the curves show the same scaling behavior with increasing $T$. In panel a, $\sigma_{AL}(T)$ is plotted in double logarithmic scale: For $\Delta n_{2D} \gtrsim 5\cdot10^{13}$ e$^{+}$/cm$^{2}$ (five uppermost curves) at any $T\gtrsim 8$ K, $\sigma_{AL}$ scales well as a power law:
\begin{equation}
\sigma_{AL}\propto T^{\beta}
\label{eq:power_law}
\end{equation} 
As in the case of the bulk curves in Fig.\ref{fig:bulk_transport}b, this scaling indicates that -- although $\sigma_{AL}$ increases with $T$ -- the gate-induced AL is not actually insulating. However, i) $\sigma_{AL}$ does not saturate below $8$ K, rapidly decreasing toward zero instead; and ii) its scaling exponent $\beta$ is always larger than $1/3$. These findings indicate that the AL does not behave as a disordered metal either, and is instead in the QC regime slightly on the insulating side of the MIT \cite{McMillanPRB1981,LarkinJETP1982,HeegerPS2002}.

On the other hand, for $\Delta n_{2D} \lesssim 5\cdot10^{13}$ e$^{+}$/cm$^{2}$ (three lowermost curves), $\sigma_{AL}$ decreases for decreasing $T$ faster than a power law in the entire $T$ range, indicating that the AL is outside the QC regime and charge transport occurs through a hopping mechanism between localized states \cite{HeegerPS2002,MottBook1979,MottBook1990}. This can be shown explicitly by plotting $\sigma_{AL}$ in semilogarithmic scale vs. $T^{-1/x}$, where $x$ can assume different values depending on the specific hopping process and the dimensionality of the system. As we show in Fig.\ref{fig:surface_transport}b, at low doping $\sigma_{AL}$ scales well with an exponent $x=3$, the hallmark of Mott variable range hopping (VRH) in $2$D \cite{HeegerPS2002,MottBook1979,MottBook1990}:
\begin{equation}
\sigma_{AL} \propto \mathrm{exp}(-B/T^{1/3})
\label{eq:exponential}
\end{equation}
This behavior is reasonable, since the field-induced carriers are strongly confined at the surface in a quasi-2D hole system, and we don't expect Coulomb interactions to play a strong role in this range of $T$ and doping \cite{BustarretPSS2008}. Indeed, $\sigma_{AL}$ does not show a good agreement with other values of $x$, such as that of Mott VRH in $3$D ($x=4$) or Efros-Shklovskii VRH in presence of a Coulomb gap ($x=2$). 


Our results show that employing B-doped diamond films increases the maximum achievable gate-induced surface carrier densities up to $\sim 3$ times with respect to undoped films as reported in the literature \cite{DankerlPRL2011,DankerlAPL2012,YamaguchiJPSJ2013,TakahidePRB2014,TakahidePRB2016,AkhgarNanoLett2016}. This likely stems from an increase in the EDL capacitance due to the sizable concentration of mobile carriers already at $V_G=0$. Despite this advantage, the sheet conductance of the AL in our NCD films remains up to $\sim 10$ times smaller than the ones reported in undoped epitaxial films and single-crystals upon field-effect doping in the entire $T$ range \cite{YamaguchiJPSJ2013,TakahidePRB2014,TakahidePRB2016,AkhgarNanoLett2016}, and the AL is unable to cross to the metallic side of the MIT even for the largest values of $\Delta n_{2D}$. This can not be simply attributed to the presence of grain boundaries, since the bulk conductivity is firmly in the disordered metal regime. However, the gate-induced AL is confined within few atomic layers \cite{PiattiPRB2017,UmmarinoPRB2017,YeScience2015,SaitoScience2015,FeteAPL2016,PiattiApsusc2018} from a surface that features a much larger roughness $\gtrsim 30$ nm, and local ``valleys'' between the grains (up to $\sim 100$ nm deep, see Fig.\ref{fig:fabrication}a) may also lead to local percolative transport through the lighly-doped bulk. Therefore, disorder may be strongly amplified at the surface were the AL is confined, shifting the metallic edge of its MIT to much larger doping values with respect to the bulk.

\begin{figure}
\begin{center}
\includegraphics[keepaspectratio,width=\columnwidth]{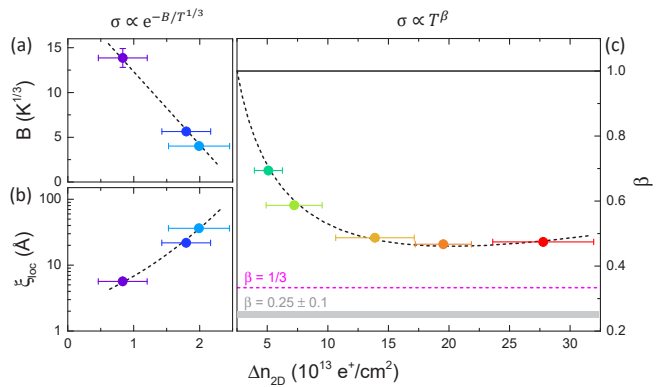}
\end{center}
\caption{
(a) Scaling factor $B$, (b) localization length $\xi_{loc}$, and (c) scaling factor $\beta$, vs. increasing hole density $\Delta n_{2D}$. Points are color-coded to match the corresponding curves in Fig.\ref{fig:surface_transport}. Black dashed lines act as guides to the eye. The dashed magenta line marks $\beta = 1/3$. The shaded grey area represents the range of $\beta$ values determined from the bulk conductivity below $20$ K.
}
\label{fig:scaling} 
\end{figure}

We can gain further insight into the surface transport close to the MIT by considering the dependence of the scaling factors in Eqs.\ref{eq:power_law} and \ref{eq:exponential}, $\beta$ and $B$, with increasing hole density. In the VRH regime, $B$ strongly decreases at the increase of $\Delta n_{2D}$ (see Fig.\ref{fig:scaling}a), and is related to the localization length $\xi_{loc}$ as \cite{MottBook1979,MottBook1990}:
\begin{equation}
B=3/k_BN(E_F)\xi_{loc}^2
\label{eq:hopping_distance}
\end{equation}
where $k_B$ is the Boltzmann constant, $N(E_F)=m^*/\pi\hbar^2$ is the 2D-DOS at the Fermi level for parabolic bands, $m^*$ is the effective mass and $\hbar$ is the reduced Planck constant. We determine $\xi_{loc}$ (see Fig.\ref{fig:scaling}b) by assuming $m^*\simeq0.8m_0$, where $m_0$ is the electron mass, which has been found to describe the DOS of diamond with a good degree of approximation \cite{IoffeInstDiam,CollinsJPC1971,ProsserCJP1964}. As expected \cite{BustarretPSS2008}, $\xi_{loc}$ is comparable with the Bohr radius of B-doped diamond  ($\sim 3.5$ \AA) at low doping, and strongly increases as the AL approaches the MIT, where it would then diverge \cite{BustarretPSS2008}. If we perform a linear extrapolation of $\xi_{loc}$ to $\Delta n_{2D}\simeq 5\cdot 10^{13}$ e$^+$/cm$^2$, we find $\xi_{loc}\sim30$ nm, comparable with the surface roughness $S_q$. Therefore, the crossover from the VRH to the QC regime occurs when $\xi_{loc}$ approaches the length scale describing the non-ideality of the surface.

In the QC regime, the behavior of the AL is described by the scaling factor $\beta$. Theory predicts \cite{McMillanPRB1981,LarkinJETP1982,HeegerPS2002} that $\beta=1$ marks the transition from the VRH to the QC regime, $1/3<\beta<1$ the insulating side of the QC regime, and $\beta<1/3$ its metallic side. $\beta$ is exactly equal to $1/3$ only at the QC point of the MIT, where Eq.\ref{eq:power_law} holds down to $T=0$ and the Fermi level is exactly at the mobility edge. In Fig.\ref{fig:scaling}c, we show that the AL sits firmly in the insulating side of the QC regime for any value of $\Delta n_{2D}$. As $\Delta n_{2D}$ increases, $\beta$ first decreases and then saturates around $\sim 0.44$ for $\Delta n_{2D}\gtrsim 1.5\cdot10^{14}$ e$^+$/cm$^2$: the AL is unable to reach the QC point (dashed magenta line) and cross into the disordered metal regime of the underlying bulk (shaded grey area). We ascribe this behavior to extra scattering centers introduced by the ions in the EDL at the solid/electrolyte interface \cite{BragaNanoLett2012,SaitoACSNano2015,PiattiApsusc2017,Gonnelli2DMater2017,PiattiAPL2017}, an effect that can even lead to re-entrant MITs in extremely surface-sensitive materials \cite{OvchinnikovNatCommun2016,LuPNAS2017}. The gate-induced disorder shifts the Anderson transition to larger doping levels, eventually compensating the concomitant increase in $\Delta n_{2D}$ and locking the AL in the QC regime on the insulating side of the MIT.

In summary, we have performed ionic gating experiments on nanocrystalline B-doped diamond films. We have shown that the presence of B doping enhances the maximum induced carrier density with respect to undoped films and single crystals. However, no sign of SC was observed down to $\sim 3$ K. By disentangling the sheet conductance of the field-induced accumulation layer from that of the underlying bulk, we were able to probe the surface transport properties as a function of temperature and induced charge density. By increasing the hole density, we observed a transition from the variable-range hopping to the quantum critical regime in the accumulation layer. Finally, we showed that the insulator-to-metal transition is never reached at the surface, and suggested that this frustrated behavior may be due to an increased disorder arising from a combination of non-negligible surface roughness and extra scattering centers introduced by the ionic gate.

\section*{Acknowledgments}
We are very thankful to D. Romanin for theoretical and experimental support, to \mbox{C. Gerbaldi} and J. R. Nair for the design and preparation of the electrolytes, and to P. Olivero and F. Laviano for scientific support. G.P.'s work at Ulm University was supported by the Erasmus Traineeship programme of the European Union. 

\section*{Author contributions}
R.S.G. and E.P. designed the experiments. E.P. and F.G. realized the EDLT devices, performed the transport measurements and analyzed the data. G.P. and A.P. grew the samples. R.S.G. performed the AFM measurements and supervised both the experiments and the data analysis. E.P., F.G. and R.S.G. wrote the manuscript with input from all authors.


\begin{thebibliography}{99}
%
\bibitem{PanBook1995} L. S. Pan and D. R. Kania, \textit{Diamond: Electronic Properties and Applications} (Kluwer Academic Publishers, 1995)
%
\bibitem{EkimovNature2004} E. A. Ekimov et al., Nature \textbf{428}, 542 (2004)
%
\bibitem{BustarretPSS2008} E. Bustarret, Phys. Status Solidi A \textbf{205}, 997 (2008)
%
\bibitem{TakanoAPL2004} Y. Takano et al., Appl. Phys. Lett. \textbf{85}, 2851 (2004)
%
\bibitem{OkazakiAPL2015} H. Okazaki et al., Appl. Phys. Lett. \textbf{106}, 052601 (2015) 
%
\bibitem{BoeriPRL2004} L. Boeri, J. Kortus, and O. K. Andersen, Phys. Rev. Lett. \textbf{93}, 237002 (2004)
%
\bibitem{LeePRL2004} K.-W. Lee and W. E. Pickett, Phys. Rev. Lett. \textbf{93}, 237003 (2004)
%
\bibitem{BoeriJPCS2005} L. Boeri, J. Kortus, and O. K. Andersen, J. Phys. Chem. Solids \textbf{67}, 552 (2006)
%
\bibitem{GiustinoPRL2007} F. Giustino et al., Phys. Rev. Lett. \textbf{98}, 047005 (2007)
%
\bibitem{NakamuraPRB2013} K. Nakamura et al., Phys. Rev. B \textbf{87}, 214506 (2013)
%
\bibitem{UenoReview2014} K. Ueno et al., J. Phys. Soc. Jpn. \textbf{83}, 032001 (2014)
%
\bibitem{DankerlPRL2011} M. Dankerl et al., Phys. Rev. Lett \textbf{106}, 196103 (2011) 
%
\bibitem{DankerlAPL2012} M. Dankerl et al., Appl. Phys. Lett. \textbf{100}, 023510  (2012)
%
\bibitem{YamaguchiJPSJ2013} T. Yamaguchi et al., J. Phys. Soc. Jpn. \textbf{82}, 074718  (2013)
%
\bibitem{TakahidePRB2014} Y. Takahide et al., Phys. Rev. B \textbf{89}, 235304  (2014)
%
\bibitem{TakahidePRB2016} Y. Takahide et al., Phys. Rev. B \textbf{94}, 161301(R) (2016)
%
\bibitem{AkhgarNanoLett2016} G. Akhgar et al., Nano Lett. \textbf{16}, 3768 (2016)
%
\bibitem{DagheroPRL2012} D. Daghero et al., Phys. Rev. Lett. \textbf{108}, 066807  (2012)
%
\bibitem{TortelloApsusc2013} M. Tortello et al., Appl. Surf. Sci. \textbf{269} 17 (2013) 
%
\bibitem{PiattiPRB2017} E. Piatti et al., Phys. Rev. B \textbf{95}, 140501(R) (2017) 
%
\bibitem{UmmarinoPRB2017} G. A. Ummarino et al., Phys. Rev. B \textbf{96}, 064509 (2017)
%
\bibitem{PiattiArXiv2018} E. Piatti et al., Phys. Rev. Materials \textbf{3}, 044801 (2019)
%
\bibitem{GonnelliSciRep2015} R. S. Gonnelli et al., Sci. Rep. \textbf{5}, 9554 (2015) 
%
\bibitem{PiattiJSNM2016} E. Piatti et al., J. Supercond. Nov. Magn. \textbf{29}, 587 (2016)
%
\bibitem{LiNature2016} L. J. Li et al., Nature \textbf{529}, 185 (2016)
%
\bibitem{PippionePSS2017} G. Pippione et al., Phys. Status Solidi A \textbf{214}, 1700223 (2017)
%
%
\bibitem{UshizawaDRM1998} K. Ushizawa et al., Diam. Relat. Mater. \textbf{7}, 1719 (1998)
%
\bibitem{InzeltBook} G. Inzelt, in \textit{Electroanalytical Methods: Guide to Experiments and Applications,} edited by F. Scholz (Springer-Verlag, Berlin, 2010), p. 147-158
%
%
\bibitem{McMillanPRB1981} W. L. McMillan, Phys. Rev. B \textbf{24}, 2739 (1981)
%
\bibitem{LarkinJETP1982} A. I. Larkin and D. E. Khmel'nitskii, J. Exp. Theor. Phys. \textbf{56}, 647 (1982)
%
\bibitem{HeegerPS2002} A. J. Heeger, Phys. Scr. \textbf{2002}, 30 (2002)
%
\bibitem{GononJAP1995} P. Gonon et al., J. Appl. Phys. \textbf{78}, 7059 (1995)
%
\bibitem{KounoArXiv2018} S. Kouno et al., Sci Rep. \textbf{8}, 14731 (2018)
%
\bibitem{MottBook1979} N. F. Mott and E. A. Davis, \textit{Electronic Processes in Noncrystalline Materials} (Oxford University Press, Oxford, 1979)
%
\bibitem{MottBook1990} N. F. Mott, \textit{Metal-Insulator Transition} (Taylor \& Francis, London, 1990)
%
\bibitem{YeScience2015} J. M. Lu et al., Science \textbf{350}, 1353 (2015)
%
\bibitem{SaitoScience2015} Y. Saito, Y. Kasahara, J. T. Ye, Y. Iwasa, and T. Nojima, Science \textbf{350}, 409 (2015)
%
\bibitem{FeteAPL2016} A. F{\^e}te, L. Rossi, A. Augieri, and C. Senatore, Appl. Phys. Lett. \textbf{109}, 192601 (2016)
%
\bibitem{PiattiApsusc2018} E. Piatti et al., Appl. Surf. Sci \textbf{461}, 17 (2018)
%
\bibitem{IoffeInstDiam} http://www.ioffe.ru/SVA/NSM/Semicond/Diamond/
bandstr.html\#masses
%
\bibitem{CollinsJPC1971} A. T. Collins and A. W. S. Williams, J. Phys. C.: Solid State Phys. \textbf{4}, 1789 (1971)
%
\bibitem{ProsserCJP1964} V. Prosser, Czech. J. Phys. \textbf{B15}, 128 (1964)
%
\bibitem{BragaNanoLett2012} D. Braga et al., Nano Lett. \textbf{12}, 5218 (2012)
%
\bibitem{SaitoACSNano2015} Y. Saito and Y. Iwasa, ACS Nano \textbf{9}, 3192 (2015) 
%
\bibitem{PiattiApsusc2017} E. Piatti et al., Appl. Surf. Sci. \textbf{395}, 37 (2017)
%
\bibitem{Gonnelli2DMater2017} R. S. Gonnelli et al., 2D Mater. \textbf{4}, 035006 (2017)
%
\bibitem{PiattiAPL2017} E. Piatti, Q. H. Chen, and J. T. Ye, Appl. Phys. Lett. \textbf{111}, 013106 (2017)
%
\bibitem{OvchinnikovNatCommun2016} D. Ovchinnikov et al., Nat. Commun. \textbf{7}, 12391 (2016)
%
\bibitem{LuPNAS2017} J. M. Lu et al., Proc. Nat. Acad. Sci. USA \textbf{115}, 3551 (2018)
%
\end{thebibliography}
\end{document}